# Modeling Headway in Heterogeneous and Mixed Traffic Flow: A Statistical Distribution Based on a General Exponential Function

**Natchaphon Leungbootnak**
Zachry Department of Civil & Environmental Engineering
Texas A&M University, 3136 TAMU, College Station, TX, USA
Email: natchaphonl@tamu.edu

**Zihao Li**
Zachry Department of Civil & Environmental Engineering
Texas A&M University, 3136 TAMU, College Station, TX, USA
Email: scottlzh@tamu.edu

**Zihang Wei**
Zachry Department of Civil & Environmental Engineering
Texas A&M University, 3136 TAMU, College Station, TX, USA
Email: wzh96@tamu.edu

**Dominique Lord**
Zachry Department of Civil & Environmental Engineering
Texas A&M University, 3136 TAMU, College Station, TX, USA
Email: d-lord@tamu.edu

**Yunlong Zhang***
Zachry Department of Civil & Environmental Engineering
Texas A&M University, 3136 TAMU, College Station, TX, USA
Email: yzhang@civil.tamu.edu

**ABSTRACT**

The ability of existing headway distributions to accurately reflect the diverse behaviors and characteristics in heterogeneous traffic (different types of vehicles) and mixed traffic (human-driven vehicles with autonomous vehicles) is limited, leading to unsatisfactory goodness of fit. To address these issues, we modified the exponential function to obtain a novel headway distribution. Rather than employing Euler's number ($e$) as the base of the exponential function, we utilized a real number base to provide greater flexibility in modeling the observed headway. However, the proposed is not a probability function. We normalize it to calculate the probability and derive the closed-form equation. In this study, we utilized a comprehensive experiment with five open datasets: highD, exiD, NGSIM, Waymo, and Lyft to evaluate the performance of the proposed distribution and compared its performance with six existing distributions under mixed and heterogeneous traffic flow. The results revealed that the proposed distribution not only captures the fundamental characteristics of headway distribution but also provides physically meaningful parameters that describe the distribution shape of observed headways. Under heterogeneous flow on highways (i.e., uninterrupted traffic flow), the proposed distribution outperforms other candidate distributions. Under urban road conditions (i.e., interrupted traffic flow), including heterogeneous and mixed traffic, the proposed distribution still achieves decent results.

# 1. INTRODUCTION

Headway, the time interval between successive vehicles traveling in the same lane, is a key microscopic traffic variable in traffic flow theory and transportation applications (*1*). It provides crucial insights into traffic dynamics, congestion levels, bottlenecks, and the effectiveness of management strategies, including signal control (*2*, *3*). Headway varies due to factors such as traffic volume levels, driving speeds, road conditions, and driver characteristics, leading to a non-typical distribution that is challenging to describe using traditional statistical distributions.

Heterogeneous traffic that contains various types of vehicles, such as trucks, introduces additional complexity to headway distribution. Existing studies have proposed different distributions to handle the differences on a case-by-case basis, such as categorizing different vehicle combinations (*4*, *5*). However, a general distribution encompassing all vehicle types is essential. By incorporating the behaviors of all vehicle types into a single framework, it provides a comprehensive overview of traffic dynamics. A holistic model would enhance the accuracy of traffic assessments, improve traffic management strategies, and streamline simulations by reducing analytical complexity.

With the integration of Autonomous Vehicles (AVs) technology, including Intelligent Transportation Systems (ITS) and Adaptive Cruise Control (ACC) (*6*–*8*), AVs and Human-Driven Vehicles (HDVs) now coexist on the roads. Before achieving fully automation, mixed traffic flow, comprising both HDVs and AVs, will become the new norm (*9*, *10*). Since the car-following behaviors of AVs and HDVs are fundamentally different (*11*), it suggests that existing headway distribution models may not accurately reflect mixed traffic flow. Consequently, there is a need to develop a general distribution that effectively captures the interaction under mixed traffic flow.

In addition to heterogeneity and mixed traffic conditions arising from variations in vehicle types, traffic flow can be classified into uninterrupted and interrupted flow. Uninterrupted flow refers to traffic conditions in which vehicle movements are primarily governed by interactions among vehicles within the traffic stream without external control equipment, such as traffic signals or stop signs, that could impede continuous movement. Uninterrupted flows are typically observed on highways. In contrast, interrupted flow is characterized by the presence of controlled or uncontrolled access points that periodically disrupt vehicle progression. Such interruptions stem from regulatory elements, including traffic signals and stop affect movement. Interrupted flows are commonly observed on urban roads (*12*, *13*). Due to the distinct characteristics of uninterrupted and interrupted traffic environments, empirical evidence indicates that headways under uninterrupted flow conditions are generally shorter than those observed under interrupted flow conditions (*14*–*18*).

Various mathematical distribution models, including single and mixed distributions, have been used to capture and estimate headway in different scenarios, as detailed in the literature review. Well-known distributions such as the shifted exponential often fail to achieve the desired finesse and pass the goodness-of-fit tests. While combined and mixed models improve headway distribution representation (*19*–*21*), they typically require substantial computational effort to estimate optimal parameters, particularly in Bayesian models.

Therefore, this study introduces a novel headway distribution model designed to address the complexity of real-world traffic, including heterogeneous traffic, mixed traffic, uninterrupted flow, and interrupted flow. The modification of the general exponential function with parameters estimated via Markov Chain Monte Carlo (MCMC) is utilized to approximate the probability. Rather than using Euler's number as the base of the exponent, as in many existing models, we applied a flexible base to better fit observed headway. The proposed unified headway distribution improves upon traditional distributions to provide enhanced accuracy and flexibility in modeling and understanding traffic behaviors, thereby contributing to more efficient traffic management and control strategies.

# 2. LITERATURE REVIEW

The research regarding vehicle headway distribution can be first dated back to early foundational studies, where Adams (*22*) studied the lengths of time elapsing between two consecutive vehicles are distributed according to the exponential law of intervals including the Gaussian distribution and negative exponential distribution. The development of vehicle headway distribution can be grouped into three distinct eras (*1*). The first era is marker by pioneering efforts in manually collected traffic data. In this era, most of the vehicle headway data are collected manually and these studies assumed the headway distribution is equivalent to the vehicle arrival rates. Moreover, since the number of vehicles operating on the road during this period was not large, all the vehicles were observed in uninterrupted traffic and all vehicles were running at free-flow speed and these studies neglected the interactions between vehicles. Nevertheless, headway samples are assumed to be independent from each other. During this period, several notable models emerged, including the shifted exponential distribution (*23*), the hyper-exponential distribution (*24*), the shifted log-normal distribution (*25*), the hyper-lang distribution (*26*), and the Pearson type III distribution (*27*).

Many other headway distribution models were proposed for more crowded roadways during the second era. In this era, more complex models were applied to fit vehicle headway distribution and moreover, the vehicle data collection technology had been largely improved. Specifically, Cowan (*28*) stated that the vehicle headway contains two components: tracking component ($V$) and free component ($U$). Cowan proposed 4 different headway distribution models (i.e., M1, M2, M3, and M4 models) with different choices of $V$ and $U$. These models have increased generality and the author concluded that the M3 and M4 models are more realistic and in most of cases, the M3 model is adequate enough. Later, Branston examined one simple model (i.e., Log-normal distribution) and two mixed models (i.e., queueing model and semi-Poisson model) to fit vehicle headway data collected through the time-lapse photography technique. The studies emerged within this era still observed all vehicles running under uninterpreted traffic situations with free-flow speed while they did not neglect the interactions between vehicles. Moreover, these studies assumed not all headway samples are independent of each other and some headways may have exact patterns. Nevertheless, some of the headway distribution models in this era assumed that vehicle headways are not conditioned on speed; however, later studies had unveiled the fact that headway distributions are dependent on speed.

The third era is characterized by the consideration of headway distributions to be speed-dependent especially when the vehicles are operating in congested flow. After examining the headway and spacing data from the Next Generation Simulation (NGSIM) Trajectory Data, Chen et al. (*29*) clearly demonstrated that the vehicle headway and spacing distributions are different under various velocity ranges. Furthermore, apart from the speed dependency property of vehicle headway distributions, they are also dependent on vehicle types. In one study by Hoogendoorn and Bovy (*30*), it has been shown that the headway distribution models can be calibrated based on specific vehicle types. Nevertheless, further studies on this topic have concluded that vehicle headway, speed, and spacing can be modeled using a joint probability distribution model. In this sense, Zou et al. (*19*) constructed a bivariate distribution model to jointly fit vehicle headway and speed data. Given the complexity of traffic behavior, existing headway distribution models typically construct various independent distributions to describe specific traffic conditions, taking into account factors such as vehicle types and traffic circumstances (*5*, *30–32*). However, studies have demonstrated that a single distribution model cannot well describe vehicle headway distribution under various roadway geometrics and traffic state conditions (*31*, *33–36*). For example, Roy and Saha (*35*) examined the various headway distribution models of two-lane roads under mixed traffic conditions and discovered that the log-logistic model is well-suited for moderate traffic flow whereas the Pearson 5 model is more accurate for congested flow. In a similar study, Singh et al. (*34*) found that the best-fitted headway distribution models can change when the traffic flow rates vary. In another study by Yin et al. (*31*), they concluded that under free-flow traffic, the log-normal distribution is suitable for fitting headway distribution while under congested traffic, the log-logistic model is more suitable. Furthermore, Wang et al. (*36*) examined the headway distribution models for different vehicle combinations and found that the headway distributions for car following car and car following truck can be modeled as log-logistic distributions while gamma distributions fit better for truck following car and truck following truck.

However, from the application point of view, these models are limited to provide a unified tool that is feasible under different vehicle combination/pair. As a remedy, there are studies tried to leverage advanced statistical approaches to develop a single distribution model. For instance, Wu et al. (*20*) applied the Bayesian Model Averaging (BMA) approach to combine the advantages of different existing distribution models. Their results demonstrate that there is not a single distribution model that can accurately describe all time headway datasets. However, by considering the advantages of different distribution models, the BMA approach is able to accurately describe headway dataset under various traffic conditions. Although efforts have been made to find a single distribution model, these advanced methods did not truly propose a unified distribution model apart from combining the characteristics of several existing distribution models.

## 3. METHODOLOGY

This study proposes a new probability density function (PDF) distribution to model headways in heterogeneous and mixed traffic. We compared it with existing headway distribution models. We applied Kolmogorov–Smirnov (KS) and Chi-square tests to examine whether each distribution model statistically aligned well with real-world data. We then employed Kullback–Leibler (KL) divergence and Wasserstein distances to quantify the differences between the model results and the observed data.

### 3.1 Propose Headway Distribution with Derivation

The proposed headway distribution consists of central tendency and dispersion parameters to model headway probability, as shown in Eq. (1). The parameter $a$ represents the measure of central tendency, aligning with the most frequent headway values observed in traffic. The parameter $b$ reflects the degree of dispersion or variability in headway observation. A lower parameter $b$ value engenders a more concentrated distribution, which indicates that the data points are closer to the central tendency represented by the parameter $a$. When parameter $b$ approaches 1, the proposed distribution resembles a uniform distribution. Since direct computation of these parameters is not feasible, MCMC is employed to derive estimates by iteratively sampling from the posterior distribution (*20*).

The proposed distribution combines exponentiation and absolute concepts with the parameters $a$ and $b$ to build up the proposed function in Eq. (1). $f(t)$ is a distance function of headway $t$, designed such that the distance value decreases as $t$ moves further away from the parameter $a$. The use of absolute captures the idea that the distance equally decreases whether the headway $t$ deviates to the left or right of the parameter $a$. It ensures that the distance function attains the maximum value of 1 when headway $t$ is equal to the parameter $a$, and decreases as the headway $t$ diverges from the parameter $a$ in either direction. However, the proposed function is not a probability function. We normalize to calculate the probability from the area of the proposed function, dividing all under the curve area, as illustrated in **Figure 1**. The function of this distribution is given in Eq. (1). Note that the detailed proof is provided below.

$$f(t) = b^{|t-a|} \tag{1}$$

Where:
      $a$ = central tendency aligning with the most frequent values
      $b$ = degree of dispersion or spread of the distribution; $0 < b < 1$

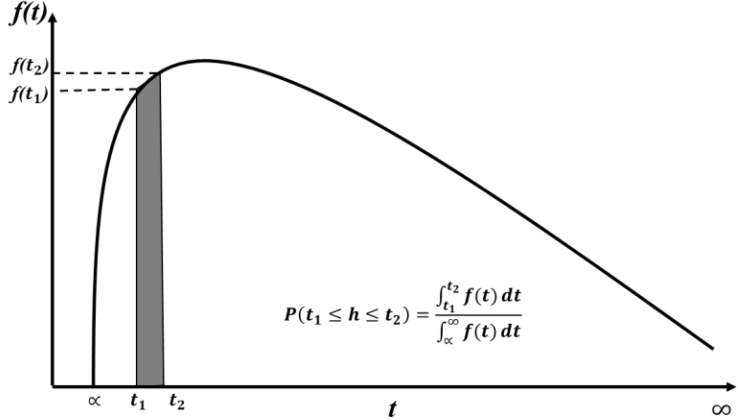

**Figure 1**. Normalization

The close form of headway probability between $t_1$ and $t_2$ from the proposed distribution is given in Eq. (2).

$$P(t_1 \leq h \leq t_2) = \begin{cases} \dfrac{b^{a-t_1}+b^{t_2-a}-2}{b^{a-\alpha}-2}; & \alpha \leq t_1 < a < t_2 \\ \dfrac{b^{a-t_1}-b^{a-t_2}}{b^{a-\alpha}-2}; & \alpha \leq t_1, t_2 \leq a \\ \dfrac{b^{t_2-a}-b^{t_1-a}}{b^{a-\alpha}-2}; & \alpha \leq a \leq t_1, t_2 \\ \dfrac{b^{t_2-a}-b^{t_1-a}}{-b^{\alpha-a}}; & a < \alpha \leq t_1, t_2 \end{cases} \quad (2)$$

Where:
$\quad$ $h$ = headway [sec]
$\quad$ $t_1, t_2$ = headway [sec]; given parameter
$\quad$ $\alpha$ = minimum headway [sec] and default = 0.5 sec

**Remark:** The parameters $t_1$ and $t_2$ satisfy $t_1, t_2 \geq \alpha$. By definition, $P(h < \alpha) = 0$, as headways shorter than the minimum threshold are considered infeasible. The minimum headway $\alpha$ is typically set to 0.5 sec, which corresponds to the lower bound of human reaction time under normal driving conditions (*37, 38*).

To better understand the derivation of the closed-form probability in Eq. (2), we provide the process of calculating the probability by applying logarithm and integration concepts to convert headway information to probability.

**Proposition 1.** The probability between $t_1$ and $t_2$ from the proposed distribution:

$$P(t_1 \leq h \leq t_2) = \frac{\int_{t_1}^{t_2} f(t)dt}{\int_{\alpha}^{\infty} f(t)dt} \quad (3)$$

**Proof.** We determine the probability from the area between $t_1$ and $t_2$ dividing all under the curve area in Eq. (4). We use an integration method to calculate the area of our proposed function.

$$P(t_1 \leq h \leq t_2) = \frac{\int_{t_1}^{t_2} f(t)dt}{\int_{\alpha}^{\infty} f(t)dt} \quad (4\text{-a})$$

$$= \frac{\int_{t_1}^{t_2} b^{|t-a|} dt}{\int_{\alpha}^{\infty} b^{|t-a|} dt} \tag{4-b}$$

We utilize the absolute theory in Eq. (5) to divide $\int_{t_1}^{t_2} b^{|t-a|} dt$ in Eq. (4-b) into two parts of the proposed function. The first part indicates $t$ that is equal to or less than the parameter $a$, while the second part specifies $t$ that is higher than the parameter $a$. We obtain:

$$f(t) = b^{|t-a|} = \begin{cases} b^{a-t}; & t < a \\ b^{t-a}; & t \geq a \end{cases} \tag{5}$$

$$\int b^{|t-a|} dt = \int b^{a-t} dt + \int b^{t-a} dt \tag{6}$$

We convert a base of exponent from $b$ to Euler's number ($e$) by using the logarithm concept:

$$\int b^{|t-a|} dt = \int e^{(a-t)lnb} dt + \int e^{(t-a)lnb} dt \tag{7-a}$$

$$= b^a \int e^{-tlnb} dt + \frac{\int e^{tlnb} dt}{b^a} \tag{7-b}$$

We apply a substitution rule in Eq. (8) to integrate our proposed function in Eq. (7). We obtain:

$$\int_{x_1}^{x_2} e^{mx} dx = \frac{e^{mx_2} - e^{mx_1}}{m} \tag{8}$$

$$\int_{t_1}^{t_2} b^{|t-a|} dt = b^a \int_{t_1}^{a} e^{-tlnb} dt + \frac{\int_{a}^{t_2} e^{tlnb} dt}{b^a}; \; t_1 < a < t_2 \tag{9-a}$$

$$= \frac{b^a(e^{-alnb} - e^{-t_1lnb})}{-lnb} + \frac{e^{t_2lnb} - e^{alnb}}{b^a lnb} \tag{9-b}$$

$$= \frac{1 - b^{a-t_1}}{-lnb} + \frac{b^{t_2} - b^a}{b^a lnb} \tag{9-c}$$

$$= \frac{b^{a-t_1} + b^{t_2-a} - 2}{lnb} \tag{9-d}$$

$$\int_{t_1}^{t_2} b^{|t-a|} dt = b^a \int_{t_1}^{t_2} e^{-tlnb} dt; \; t_1, t_2 \leq a \tag{10-a}$$

$$= \frac{b^a(e^{-t_2lnb} - e^{-t_1lnb})}{-lnb} \tag{10-b}$$

$$= \frac{b^{a-t_1} - b^{a-t_2}}{lnb} \tag{10-c}$$

$$\int_{t_1}^{t_2} b^{|t-a|} dt = \frac{\int_{t_1}^{t_2} e^{tlnb} dt}{b^a}; \; t_1, t_2 \geq a \tag{11-a}$$

$$= \frac{e^{t_2lnb} - e^{t_1lnb}}{b^a lnb} \tag{11-b}$$

$$= \frac{b^{t_2-a} - b^{t_1-a}}{lnb} \tag{11-c}$$

Akin to Eqs. (5-11), we do the same process to $\int_\alpha^\infty b^{|t-a|} dt$ in Eq. (4-b). We obtain:

$$\int_\alpha^\infty b^{|t-a|} dt = \frac{b^{a-\alpha} - 2}{lnb}; a > \alpha \tag{12}$$

$$\int_\alpha^\infty b^{|t-a|} dt = \frac{-b^{\alpha-a}}{lnb}; a \leq \alpha \tag{13}$$

We substitute Eqs. (9-11) into Eqs. (12) and (13) for the closed-form equation of the proposed distribution. We obtain Eq. (2).

**Remark**: Using a flexible base $b$ introduces an additional tuning parameter, allowing the model to more precisely capture varying decay or growth rates in the data compared to an exponential-based model. This approach is analogous to methods in Weibull-type distributions, which have been shown to provide improved fit in complex settings (*39*).

### 3.2 Parameter Estimation via Markov chain Monte Carlo (MCMC)

The MCMC is a computational method that generates samples from complex probability distributions by constructing a Markov chain to approximate the desired distribution. The MCMC samples from complex posterior distributions enable robust uncertainty quantification. This makes MCMC an ideal tool to estimate the parameters of the proposed distribution model due to its nonlinearity and complexity. It is important to note that choosing prior distribution is important for the estimation outcomes since prior distribution is linked to the prior knowledge about the parameters and it can affect the posterior distribution. Specifically, this study utilized the Metropolis-Hastings algorithm for each distribution by using the PyMC3 package and applying the Theano package to make computations faster (*40*). The model prior is set to equal since there is no previous knowledge regarding which model would better represent the distribution. 10000 iterations of two chains are simulated. The moving average becomes nearly constant after a certain number of iterations, although the chain values fluctuate up and down within a range. Chains are approximately converged after 5000 iterations. Therefore, the first 5000 iterations are used for warmup simulations, and the next 5000 samples are drawn for posterior distribution estimation. We can estimate the parameters by using the mean of all 5000 posterior samples (*20*).

### 3.3 Baseline Distributions

This study compares the proposed model to widely used models in headway analysis. This study aims to demonstrate their versatility and effectiveness in capturing the complexities of heterogeneous and mixed traffic. There are six candidate distributions: Shifted Log-normal, Weibull, Log-logistic, Gamma, Burr, and Shifted Exponential distributions. We employed MCMC and selected prior distributions to estimate the unknown parameters of each distribution. For the proposed distribution, a normal distribution is used for parameter $a$ because it better represents parameter $a$ that has a central tendency property. A uniform distribution is used for parameter $b$ because it can guarantee the range of parameter $b$ between 0 and 1. For other distributions, we utilized the same prior distributions as the previous study done by Wu et al. (*20*). Probability density function and prior distribution of unknown parameters in each distribution are used in the MCMC estimation, listed in **Table 1**.

**Table 1.** Propose and baseline distributions

| Distribution | Probability density function | Model parameter priors |
|---|---|---|
| Proposed | $f(t\|a,b) = \frac{b^{\|t-a\|}}{\int_\alpha^\infty b^{\|t-a\|} dt}$ | $a \sim N(0,10), b \sim uniform(0,1)$ |

| Distribution | Probability density function | Model parameter priors |
|---|---|---|
| Shifted Log-normal | $f(t\|\mu,\sigma,\gamma) = \dfrac{1}{(t-\gamma)\sigma\sqrt{2\pi}} e^{-\dfrac{[\ln(t-\gamma)-\mu]^2}{2\sigma^2}}; t > \gamma, t > 0$ | $\mu \sim N(0,10), \sigma \sim \Gamma(0.5, 0.5),$ $\gamma \sim uniform(-10, \min(t))$ |
| Weibull | $f(t\|\alpha,\beta) = \dfrac{\alpha t^{\alpha-1} e^{-\left(\dfrac{t}{\beta}\right)^\alpha}}{\beta^\alpha}; t \geq 0$ | $\alpha, \beta \sim \Gamma(0.5, 0.5)$ |
| Log-logistic | $f(t\|\alpha,\beta) = \dfrac{\dfrac{\alpha}{\beta}\left(\dfrac{t}{\beta}\right)^{\alpha-1}}{\left[1+\left(\dfrac{t}{\beta}\right)^\beta\right]^2}; t > 0$ | $\alpha, \beta \sim \Gamma(0.5, 0.5)$ |
| Gamma | $f(t\|\alpha,\beta) = \dfrac{\beta^\alpha}{\Gamma(\alpha)} t^{\alpha-1} e^{-\beta t}; t > 0$ | $\alpha, \beta \sim \Gamma(0.5, 0.5)$ |
| Burr | $f(t\|\alpha,\beta,\lambda) = \dfrac{\alpha\beta}{\lambda}\left(\dfrac{t}{\lambda}\right)^{\alpha-1}\left[1+\left(\dfrac{t}{\lambda}\right)^\alpha\right]^{-\beta-1}; t > 0$ | $\alpha, \beta, \lambda \sim \Gamma(0.5, 0.5)$ |
| Shifted Exponential | $f(t\|\lambda,\gamma) = \lambda e^{-\lambda(t-\gamma)}; t \geq \gamma, t \geq 0$ | $\lambda \sim \Gamma(0.5, 0.5),$ $\gamma \sim uniform(-10, \min(t))$ |

### 3.4 Statistical Tests

In this study, we used Kolmogorov-Smirnov (KS) and Chi-square tests to investigate whether each distribution has the same cumulative distribution function (CDF) with the field data. The KS test is a non-parametric test that determines the gaps between two CDFs, as listed in Eq. (14). KS hypotheses are set up as follows:

$H_0$ : The field data sample $Q$ and the distribution data sample $P$ are drawn from identical continuous CDFs.
$H_1$ : The field data sample $Q$ and the distribution data sample $P$ are drawn from different continuous CDFs.

$$D - statistics = \text{supremum}|Q(x) - P(x)| \qquad (14)$$

Since the KS test can examine only whether each distribution has the same CDF as the field data, the Chi-square test that can determine the difference between two distributions with the binned data is employed. Chi-square is computed by using Eq. (15). Chi-square hypotheses are set up as follows:

$H_0$ : The observed distribution is consistent with the expected distribution.
$H_1$ : The observed distribution significantly deviates from the expected distribution.

$$\chi^2 = \sum \frac{(O_i - E_i)^2}{E_i} \qquad (15)$$

Where:
$O_i$ = observed frequencies; $O_i \geq 5$; if frequencies in a bin < 5, it will be grouped with bins nearby until the frequency $\geq 5$

### 3.5 Performance Metrics

In this study, we also applied the KL divergence (*41*) and Wasserstein distance (*42*) to evaluate the difference of each distribution on the selected data samples. KL divergence quantifies the information lost when using a distribution $Q$ to approximate the actual distribution $Q$ from the sample dataset, as listed in Eq. (16). If the KL divergence is close to zero, it indicates that the two distributions are similar. As the KL divergence increases, it implies increasing dissimilarity between two distributions.

$$D_{KL}(P||Q) = \sum_{x \in X} P(x) log\left(\frac{P(x)}{Q(x)}\right) \tag{16}$$

If the two distributions $P$ and $Q$ are significantly distant and have no overlap, the KL divergence becomes inconclusive. To address this limitation, we introduce an alternative metric called the Wasserstein distance, which computes the minimum cost needed to transform one probability distribution into another, as illustrated in Eq. (17). Similar to the KL divergence, a smaller Wasserstein distance indicates a higher level of similarity between the compared distributions.

$$W_p(P, Q) = \left(\frac{1}{n}\sum_{i=1}^{n}\|X_{(i)} - Y_{(i)}\|^p\right)^{\frac{1}{p}} \tag{17}$$

## 4. HEADWAY DATASETS

We utilized car-following trajectories from five open-accessible traffic datasets: highD, exiD, Next Generation Simulation (NGSIM), Waymo, and Lyft datasets. **highD** is an extensive driving dataset that contains two vehicle types: truck and car. It includes precise location and speed information for each vehicle, which is issued by the Institute of Automotive Engineering at RWTH Aachen University in Germany (*14*, *43*). **exiD** consists of four vehicle types: truck, van, car, and motorcycle, which collects data by using the camera's record from drones at the entry and exit of many German expressways between Aachen and Cologne (*15*). **NGSIM** datasets are developed by the Federal Highway Administration (FHWA). The car-following events for I-80 are gathered using the reconstructed data (*16*, *44*).

The previously mentioned three datasets are based on human-driven vehicles (HDVs) where highD and exiD are uninterrupted flow datasets and NGSIM is an uninterrupted flow dataset. However, with the development of automated vehicles (AVs), AVs have already co-existed with HDVs on interrupted flow (*45*, *46*). To conduct a more comprehensive analysis of car-following behavior, and unlike previous studies that focused solely on HDVs, we introduce the AV car-following dataset from **Waymo**, a self-driving car company operated by Google. The Waymo dataset provides comprehensive annotations of object information, precise 3D vehicle poses, and high-resolution sensor data from LiDAR and cameras. It captures interactions between HDVs and AVs, encompassing a wide range of driving contexts. It includes 1,950 different driving scenarios, each lasting 20 seconds, covering both highways and urban roads (*17*, *43*, *47*). Additionally, 1,440 car-following events have been manually extracted from video data by (*48*). **Lyft** dataset comprises Level 5 autonomous driving data collected from a fleet of 20 AVs operating along a fixed route over a four-month period, encompassing mixed traffic flow conditions. It contains over 170,000 scenarios that capture interactions between HDVs and AVs (*18*). The details including viewpoint, roadway type, traffic flow scenario, and sampling frequency are summarized in **Table 2** (*15*, *43*).

Table 2. Three car-following datasets

| Dataset | Viewpoint | Roadway | Traffic flow | Sensors | AV involved | Frequency (HZ) |
|---|---|---|---|---|---|---|
| highD | External | Highway | Uninterrupted | Camera | No | 25 |
| exiD | External | Highway | Uninterrupted | Camera | No | 25 |
| NGSIM | External | Highway and Urban road | Interrupted | Camera | No | 10 |
| Waymo | Driver | Highway and Urban road | Interrupted | Camera and Lidar | Yes | 10 |
| Lyft | Driver | Urban road | Interrupted | Camera and Lidar | Yes | 10 |

We first standardize the sampling frequency by extracting headway data points every second, as each dataset collects data at varying frequencies. Subsequently, we refine the data by excluding headway periods less than 0.5 seconds, which are generally recognized as below the minimum safe headway threshold (*37*, *38*), and those greater than 25 seconds, as they do not represent car-following behavior (*49*).

## 4.1 Preliminary Car-Following Data Analysis

Descriptive statistics and distributions for the five datasets, namely NGSIM, Waymo, highD, exiD, and Lyft, were examined across three primary car-following variables: headway, spacing, and following speed, as illustrated in Figure 2. The preliminary analysis shows that NGSIM and Waymo exhibit shorter spacing than highD, exiD, and Lyft. In terms of speed, highD and exiD, which represent highway driving conditions, display higher velocities with two central distributions, whereas NGSIM, Waymo, and Lyft, collected in urban settings, reveal generally lower speeds. Among them, Waymo records the slowest following speeds. The maximum speed observed in the Lyft dataset is lower than that of the other datasets, consistent with its focus on urban environments where speed limits are typically more restrictive. For headway analysis, the datasets representing interrupted flow conditions, including NGSIM, Waymo, and Lyft, show longer average headways compared with highD and exiD, which correspond to uninterrupted freeway flow. Among these, Lyft demonstrates the highest average headway. The headway distributions of Waymo and Lyft, which contain mixed traffic with both human-driven and automated vehicles, appear more dispersed than those of the other datasets. This broader variation suggests that the inclusion of autonomous vehicles increases heterogeneity in car-following behavior, indicating that their integration may alter overall traffic dynamics. These findings highlight the importance of further investigation into the interactions between human-driven and automated vehicles in mixed-traffic environments.

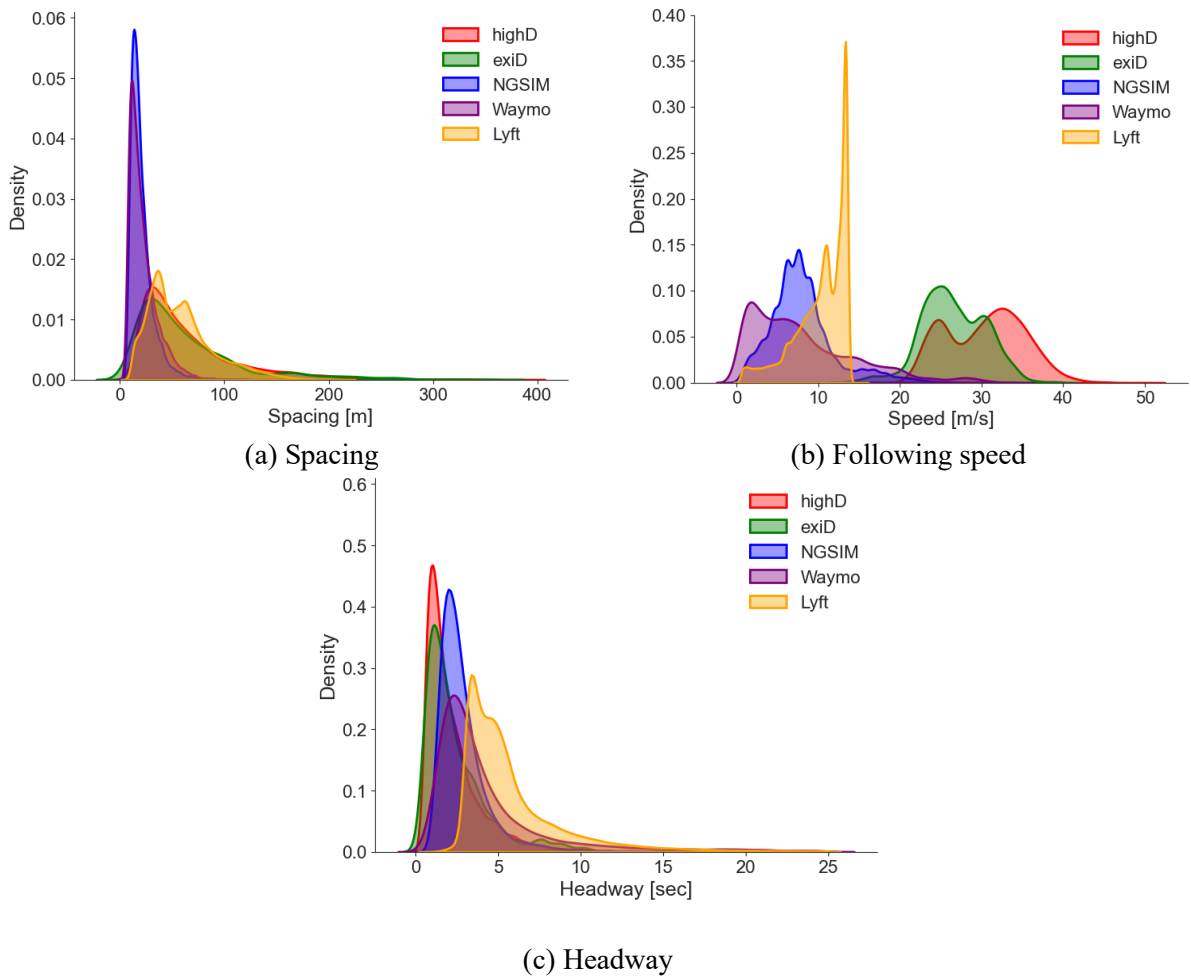

(a) Spacing

(b) Following speed

(c) Headway

**Figure 2.** Distributions of car-following measurements

|        | highD | exiD  | NGSIM | Waymo | Lyft  |
|--------|-------|-------|-------|-------|-------|
| highD  | 0     | 0.050 | 0.348 | 0.380 | 0.748 |
| exiD   |       | 0     | 0.312 | 0.349 | 0.701 |
| NGSIM  |       |       | 0     | 0.197 | 0.638 |
| Waymo  |       |       |       | 0     | 0.448 |
| Lyft   |       |       |       |       | 0     |

**Figure 3.** Pairwise comparisons of KS-statistics among headway distributions

KS-statistics between highD and exiD is the lowest. However, the KS p-values of all pairs are very close to 0. This indicates that the headway distribution in each sample does not statistically come from the same population. Although highD and exiD have the same traffic and driving scenario, their headway data are gathered from different locations. The analysis reveals that each sample has unique characteristics based on their facilities and driving scenarios.

## 5. RESULTS AND DISCUSSIONS
### 5.1 Parameter Estimations

This section presents the estimation results of the unknown parameters of the proposed distribution as well as the baseline distributions. For each distribution, the unknown parameters are estimated by MCMC with 10000 iterations, separately for each data sample. As an example, **Figure 4 (a)** illustrates the estimated posterior distribution of parameters $a$ and $b$ in the proposed distribution, fitted on the highD data sample. **Figure 4 (b)** demonstrates historical trace plots of $a$ and $b$ over 10000 iterations, where the first 5000 iterations serve as the warmup period and the remaining 5000 iterations are considered as drawn samples for the estimated posterior distributions. Parameters $a$ and $b$ in **Figure 4** can be calculated by using the mean of 5000 drawn samples with the result of 0.936 and 0.540, respectively. **Table 3** listed the MCMC estimated parameters' values for the proposed distribution and all baseline distributions, fitted on four different data samples.

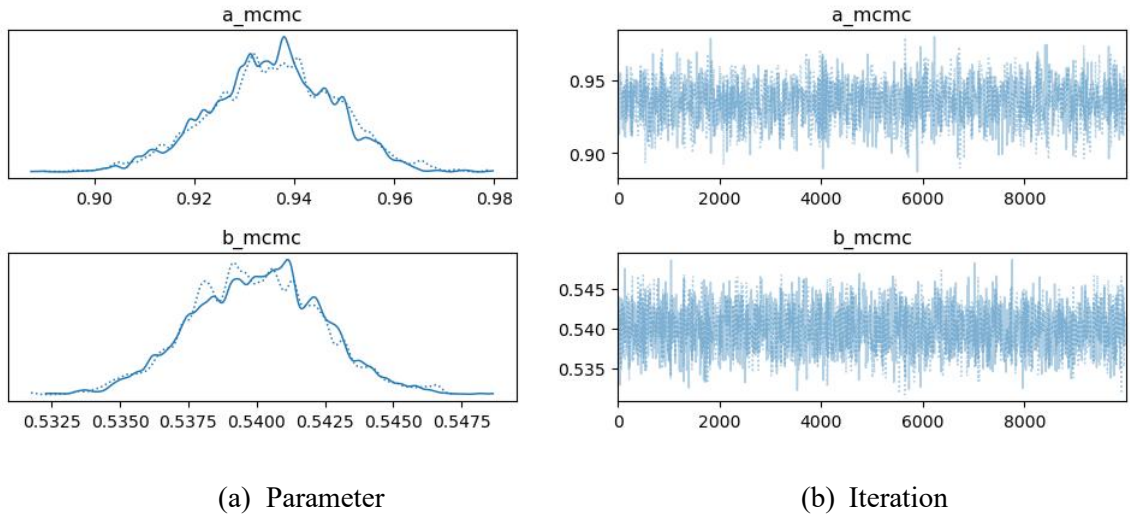

(a) Parameter  (b) Iteration

**Figure 4.** MCMC sampling plots of proposed distribution for highD

**Table 3.** Estimated parameters via MCMC

| Distribution | Parameter | highD | exiD  | NGSIM | Waymo | Lyft  |
|--------------|-----------|-------|-------|-------|-------|-------|
| Proposed     | $a$       | 0.936 | 0.879 | 2.277 | 2.339 | 4.598 |

| Distribution | Parameter | highD | exiD | NGSIM | Waymo | Lyft |
|---|---|---|---|---|---|---|
| Shifted Log-normal | $b$ | 0.540 | 0.583 | 0.481 | 0.721 | 0.676 |
| | $\mu$ | 0.233 | 0.306 | 0.683 | 1.012 | 1.448 |
| | $\sigma$ | 0.899 | 0.942 | 0.594 | 0.794 | 0.525 |
| Weibull | $\gamma$ | 0.377 | 0.374 | 0.528 | 0.483 | 0.892 |
| | $\alpha$ | 1.481 | 1.408 | 1.744 | 1.348 | 1.926 |
| | $\beta$ | 2.473 | 2.677 | 3.305 | 4.780 | 6.649 |
| Log-logistic | $\alpha$ | 2.574 | 2.419 | 3.910 | 2.686 | 4.082 |
| | $\beta$ | 1.719 | 1.826 | 2.515 | 3.215 | 5.019 |
| Gamma | $\alpha$ | 2.335 | 2.098 | 4.175 | 2.137 | 4.654 |
| | $\beta$ | 1.055 | 0.868 | 1.428 | 0.494 | 0.794 |
| Burr | $\alpha$ | 3.199 | 2.796 | 5.237 | 4.018 | 10.609 |
| | $\beta$ | 0.602 | 0.709 | 0.524 | 0.439 | 0.203 |
| | $\lambda$ | 1.296 | 1.480 | 2.021 | 2.185 | 3.387 |
| Shifted Exponential | $\lambda$ | 0.584 | 0.522 | 0.423 | 0.261 | 0.201 |
| | $\gamma$ | 0.500 | 0.499 | 0.558 | 0.506 | 0.894 |

Note: $\alpha = 0.5$ sec for all samples.

For each distribution, their estimated parameter values are different when fitted with different data samples since each sample is collected under different characteristics and environments. Moreover, the parameter values of the proposed distribution can provide meaningful interpretations. Specifically, parameter $a$ stands for the most frequent values and parameter $b$ stands for dispersion or spread. If parameter $b$ is close to 1, the frequencies are uniformly distributed. Parameter $a$ from highD and exiD are less than those in NGSIM, Waymo, and Lyft. This indicates that most headways in highD and exiD are shorter than those in NGSIM, Waymo, and Lyft. As for parameter $b$, larger parameter $b$ indicates the headway distribution is more dispersed. Only parameter $b$ estimation in NGSIM cannot represent the observation well.

## 5.2 Headway Distributions

Five samples are analyzed to evaluate the proposed and six candidate distributions using 49 bins of 0.5-second headways ranging from 0.5 to 25 seconds. We plot observation frequency along with the proposed distribution and three outstanding baseline distributions, as shown in **Figure 4**. The proposed and Shifted Log-normal graphically fit with observations in uninterrupted flow of highD and exiD. Log-logistic and Burr match field data in interrupted flow with heterogeneous traffic of NGSIM. Only Burr seems to capture observed data in an interrupted flow with mixed traffic of Waymo and Lyft. In the interrupted flow condition, our proposed distributions are more dispersed than observed distributions because parameters $b$ estimation are too low.

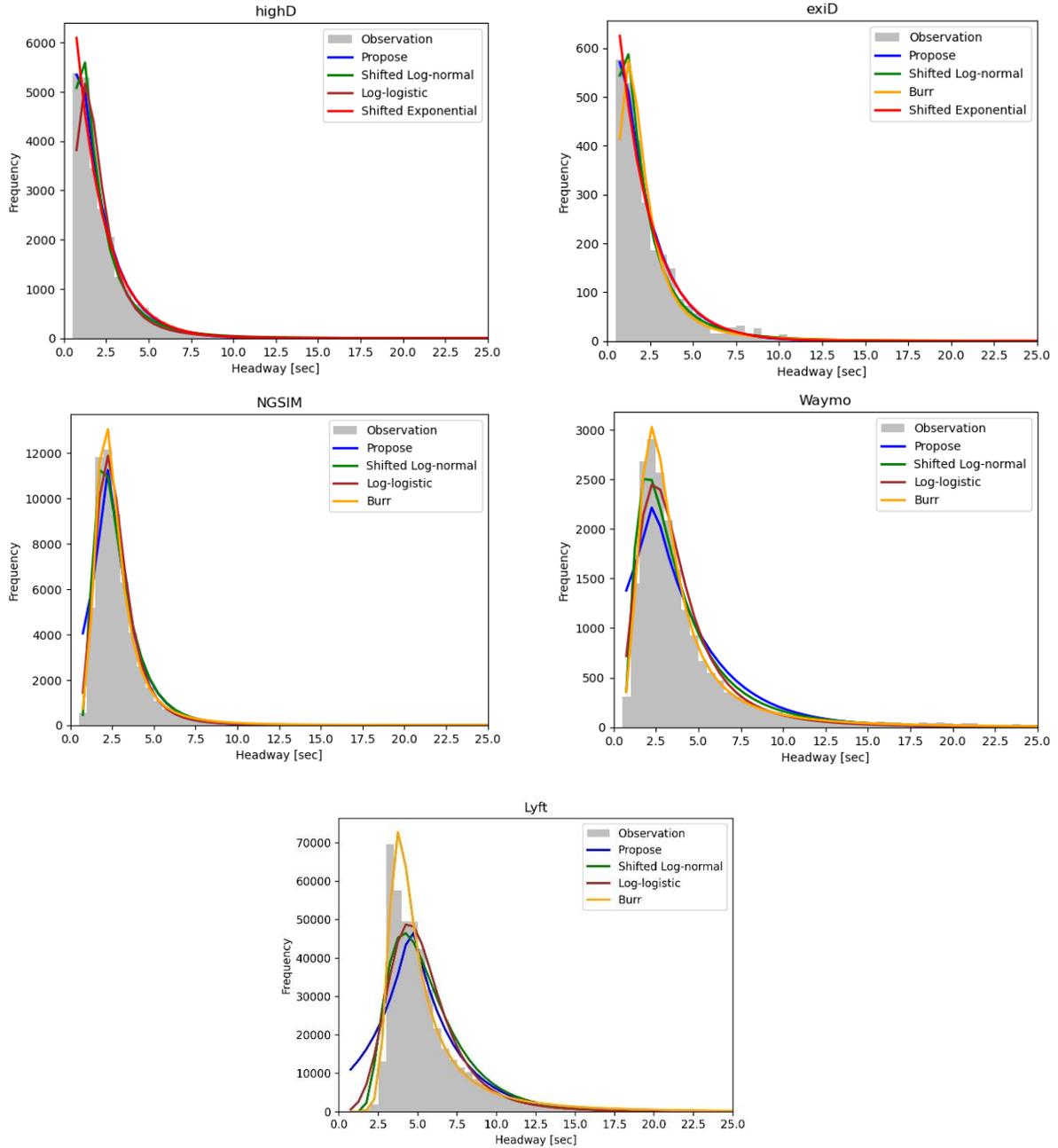

**Figure 5** Frequency histogram plot with fitted distributions

Furthermore, **Table 4** lists the p-values of Chi-square and KS test to quantify the difference between each distribution and the observed data. In **Table 5**, KL divergence and Wasserstein distance are presented, representing the goodness of fit of each distribution matching with the observed data. If the p-value exceeds 0.05, it will fail to reject $H_0$ at a confidence level of 95%. In Chi-square test, $H_0$ indicates no difference between the expected and observed distributions. While $H_0$ in KS test signifies that the expected and observed distributions have the same CDFs.

**Table 4.** Statistical tests of proposed distribution against baseline distributions

| Distribution | Chi-square p-value | | | | | Kolmogorov–Smirnov p-value | | | | |
|---|---|---|---|---|---|---|---|---|---|---|
| | highD | exiD | NGSIM | Waymo | Lyft | highD | exiD | NGSIM | Waymo | Lyft |
| Proposed | 0.0000 | 0.0000 | 0.0000 | 0.0000 | 0.0000 | 1.0000** | 1.0000** | 0.9421** | 0.9875** | 0.1865** |
| Shifted Log-normal | 0.0000 | 0.0000 | 0.0000 | 0.0000 | 0.0000 | 1.0000** | 1.0000** | 1.0000** | 1.0000** | 0.9799** |
| Weibull | 0.0000 | 0.0000 | 0.0000 | 0.0000 | 0.0000 | 0.1490** | 0.1275** | 0.2122** | 0.1293** | 0.1762** |
| Log-logistic | 0.0000 | 0.0000 | 0.0000 | 0.0000 | 0.0000 | 0.9521** | 0.9751** | 1.0000** | 1.0000** | 0.9279** |
| Gamma | 0.0000 | 0.0000 | 0.0000 | 0.0000 | 0.0000 | 0.0223 | 0.2054** | 0.0000 | 0.1186** | 0.0000 |
| Burr | 0.0000 | 0.0000 | 0.0000 | 0.0000 | 0.0000 | 0.9964** | 0.9888** | 1.0000** | 1.0000** | 1.0000** |
| Shifted Exponential | 0.0000 | 0.0000 | 0.0000 | 0.0000 | 0.0000 | 1.0000** | 1.0000** | 0.0023 | 0.2215** | 0.0000 |

Note: ** stands for fail to reject $H_0$ at a confidence level of 95%.

Chi-square results show that all fitted distributions differ statistically from the observed data across all scenarios at the 95% significance level. For the KS tests, with the exception of Gamma in highD, NGSIM, and Lyft, and Shifted Exponential in NGSIM and Lyft, all distributions exhibit the same cumulative distribution function as the observations under all conditions at the 95% significance level. Most findings from the KS analysis are consistent with previous studies (*1*, *20*, *31*, *33*). The proposed distribution demonstrates superior performance in scenarios representing uninterrupted highway flow. Among the alternative distributions, Shifted Log-normal, Log-logistic, and Burr show satisfactory performance across all conditions, whereas Weibull and Gamma do not. Shifted Log-normal performs better than Shifted Exponential, particularly under interrupted flow characterized by lower speeds and larger headways.

**Table 5.** Performance comparison of proposed distribution against baseline distributions

| Distribution | Kullback-Leibler (KL) divergence | | | | | Wasserstein distance | | | | |
|---|---|---|---|---|---|---|---|---|---|---|
| | highD | exiD | NGSIM | Waymo | Lyft | highD | exiD | NGSIM | Waymo | Lyft |
| Proposed | **0.0032** | **0.0197** | 0.0805 | 0.0854 | 0.1987 | **0.0011** | **0.0014** | 0.0040 | 0.0056 | 0.0072 |
| Shifted Log-normal | 0.0111 | 0.0277 | 0.0157 | 0.0242 | 0.0798 | 0.0016 | 0.0016 | 0.0021 | 0.0031 | 0.0056 |
| Weibull | 0.0534 | 0.0668 | 0.2098 | 0.1612 | 0.3063 | 0.0052 | 0.0054 | 0.0088 | 0.0085 | 0.0122 |
| Log-logistic | 0.0287 | 0.0440 | 0.0173 | 0.0375 | 0.1044 | 0.0026 | 0.0029 | 0.0020 | 0.0035 | 0.0048 |
| Gamma | 0.0454 | 0.0612 | 0.1182 | 0.1337 | 0.1977 | 0.0049 | 0.0048 | 0.0173 | 0.0077 | 0.0186 |
| Burr | 0.0332 | 0.0471 | **0.0027** | **0.0076** | **0.0207** | 0.0032 | 0.0025 | **0.0011** | **0.0010** | **0.0018** |
| Shifted Exponential | 0.0065 | 0.0208 | 0.3024 | 0.1594 | 0.5274 | 0.0018 | 0.0020 | 0.0047 | 0.0047 | 0.0082 |

In both KL and Wasserstein distance analyses, the proposed distribution consistently provides the best performance under uninterrupted flow with heterogeneous traffic represented by the highD and exiD datasets. It is followed by either the Shifted Log-normal or the Shifted Exponential distributions. In contrast, the Burr distribution outperforms all others in interrupted flow with heterogeneous traffic in NGSIM and in mixed traffic conditions in Waymo and Lyft, where it is followed by either the Shifted Log-normal or the Log-logistic distributions. The proposed distribution still achieves acceptable performance in these cases. Furthermore, both the Shifted Log-normal and the Log-logistic distributions show reliable performance across all situations, with the former performing slightly better. Based on the overall statistical analysis and goodness-of-fit evaluation, the proposed distribution and Burr provide the most accurate representation of headway behavior in highway-driving and urban-driving scenarios, respectively. The Shifted Log-normal maintains reasonably good performance across all conditions. The proposed distribution and the Shifted Exponential are particularly suitable for uninterrupted highway flow, while Burr and Log-logistic are more appropriate for interrupted traffic on urban roads.

Overall, the proposed distribution demonstrates superior performance under uninterrupted flow and maintains satisfactory accuracy under interrupted conditions compared with the best-performing alternatives, such as Burr. These findings suggest that it can serve as a promising and robust framework for modeling headway distributions across diverse traffic scenarios.

## 6. CONCLUSIONS

In this study, we proposed a novel statistical distribution to more accurately describe vehicle headway behavior, and its mathematical derivation was provided. The proposed distribution introduces exponentiation and absolute value operations to estimate the probability of observed headways, using a flexible base for the exponent instead of the fixed natural logarithm base. This design allows the model to better capture diverse headway patterns under different traffic conditions.

To validate its performance, we conducted numerical comprehensive experiments comparing the proposed distribution with six existing ones, including Shifted Log-normal, Weibull, Log-logistic, Gamma, Burr, and Shifted Exponential. Parameters for each model were estimated using MCMC method. Five public trajectory datasets, namely highD, exiD, NGSIM, Waymo, and Lyft, were used to represent heterogeneous and mixed traffic under both uninterrupted and interrupted flow conditions. The observed headway distributions of these datasets differ statistically from one another. In general, headways on highways with uninterrupted flow are shorter than those in urban interrupted flow, while headways in CAV traffic are more dispersed than those in HDV traffic. The Chi-square test, KS test, KL divergence, and Wasserstein distance were used to evaluate the performance of all distributions.

Results show that the parameters of the proposed model provide meaningful descriptions of the observed distribution shapes. At the 95% significance level, the Chi-square test shows that all distributions differ from the observed data, while the KS test indicates that all distributions except Gamma and Shifted Exponential closely match the cumulative distribution of the field headways. In terms of KL divergence and Wasserstein distance, the proposed distribution achieves the best performance under uninterrupted highway flow with heterogeneous traffic. Under interrupted flow with both heterogeneous and mixed traffic scenarios, the Burr distribution performs best, yet our proposed model still achieves comparable and stable results. This consistent performance across both highway and urban conditions demonstrates the generality and robustness of the proposed approach.

Although this paper demonstrates that the proposed distribution performs reliably and consistently across various traffic conditions, it still has certain limitations. The datasets used in this study do not include uninterrupted flow under mixed traffic conditions, which limits the scope of generalization. Future studies should expand the dataset coverage to address this limitation. Although MCMC provides accurate parameter estimation, it is computationally expensive. Therefore, future work will focus on using larger headway datasets and developing closed-form parameter approximations with clear physical meaning to improve computational efficiency without sacrificing accuracy.